# Probing the Galactic dark matter mass function using microlensing and direct searches

E. J. Kerins and B. J. Carr[a]

[a]Astronomy Unit, School of Mathematical Sciences, Queen Mary & Westfield College, Mile End Road, London E1 4NS, United Kingdom

If compact baryonic objects contribute significantly to the dark matter in our Galaxy, their mass function will present vital clues for galaxy formation theories and star formation processes in the early Universe. Here we discuss what one might expect to learn about the mass function of Galactic dark matter from microlensing and from direct searches in the infrared and optical wavebands. Current microlensing results from the *EROS* collaboration already constrain halo mass functions which extend below $10^{-4}$ $M_\odot$, whilst recent *HST* observations place strong constraints on disc and halo dark matter mass functions extending above 0.1 $M_\odot$. Infrared observations should either detect or constrain objects larger than 0.01 $M_\odot$ in the near future. Objects below 0.01 $M_\odot$ should be detectable through microlensing, although the prospects of determining their mass function depend critically on a number of factors.

## 1. INTRODUCTION

There is evidence for dark matter on many scales: in the Galactic disc; in galactic haloes; in clusters of galaxies; even perhaps in a smooth background [1,2]. The evidence implies that, as one observes on larger scales, the discrepancy between the amount of material inferred from an inventory of the luminous matter and that implied by dynamical studies increases, and that the luminous matter contributes only around 0.3% of the critical density [3].

The solution to these dark matter problems may not be the same for each case. Indeed, whilst any disc dark matter must surely be baryonic (since one requires dissipation for the formation of discs), the cluster and background dark matter problems almost certainly require non-baryonic solutions, since structure formation and primordial nucleosynthetic arguments favour a low baryon density (at around 5% of the critical density) [4,5]. The nature of halo dark matter remains unknown and may be baryonic, non-baryonic or a mixture of both.

For this talk we assume that baryonic dark matter (BDM) contributes significantly to the density in galactic haloes. BDM represents the minimal hypothesis in that no new fundamental physics is required: we know that baryons exist! Also, the amount of matter inferred to be locked up in haloes is intriguingly close to the prediction of primordial nucleosynthesis calculations. Therefore, the BDM hypothesis seems a reasonable one. BDM candidates span 13 orders of magnitude from $10^{-7}$-$M_\odot$ 'snowballs' to $10^6$-$M_\odot$ supermassive black holes, although there are observational and theoretical constraints on many of these [2]. If the dark matter in galaxies is in the form of compact baryonic objects it is clearly desirable to know their mass function. The mass function is crucial to understanding processes involved in star and galaxy formation and so is not only relevant to the dark matter problem.

## 2. THE MASS FUNCTION

The mass function (MF) is the number density per unit mass interval, $dn/dm$, of objects in the mass range $(m, m+dm)$ and is typically expressed in some functional form, such as a power law:

$$dn/dm \propto m^{-\gamma} \qquad (m_l < m < m_u), \qquad (1)$$

where $m_l$ and $m_u$ denote the lower and upper mass cutoffs, respectively. The problem in deciding a functional form for the dark matter MF is that current star formation theories are not suffi-ciently advanced to reliably predict it. Therefore one is forced to adopt the simplest case and so we shall assume that the dark matter MF has the form of eqn (1). Note from eqn (1) that the mass density $\rho \propto m^{2-\gamma}$ and so $\rho$ is dominated by objects of mass $m_l$ if $\gamma > 2$ and by objects of mass $m_u$ if $\gamma < 2$.

Observations of visible disc stars imply that $\gamma$ decreases from 2.7, between 1 $M_\odot$ and 10 $M_\odot$ [6], to around 1.3, between 0.1 $M_\odot$ and 0.5 $M_\odot$ [7], implying that low mass M-dwarfs and brown dwarfs (stars lighter than 0.08 $M_\odot$, which have insufficient mass for hydrogen burning) may not contribute greatly to the disc dark matter density. Some observations of stars in the visible spheroid imply a very steep MF with $\gamma \simeq 3.5$ down to 0.15 $M_\odot$ [8], although recent *HST* data appears to contradict these claims [9]. The halo MF is completely unknown. The different values of $\gamma$ for the disc and spheroid highlight a further problem in reconstructing the dark matter MF from observations. The fact that the Galaxy is characterised by several distinct components implies that star formation conditions in these components are likely to have been quite different from each other and so one should not expect their MFs to be the same. Therefore one needs to be able to disentangle observations of disc dark matter from halo dark matter in order to specify $\gamma$ for each. In the case of infrared or optical searches this is possible with additional spectroscopic and velocity dispersion data but, as we shall discuss shortly, this is a non-trivial matter for microlensing.

## 3. MICROLENSING SEARCHES

The microlensing technique (which is described elsewhere in these proceedings [10]) was first recognised by Paczyński as a potentially useful method for detecting dark matter in our Galaxy [11]. There are presently three microlensing experiments in progress: *MACHO* [12], which has so far been monitoring stars towards the Large Magellanic Cloud and Galactic Bulge; *EROS* [13], which so far has only observed the LMC, and *OGLE* [14], which is looking at the Galactic Bulge. Hitherto 6 events have been detected by *MACHO* and *EROS* towards the LMC and more than 50 by *MACHO* and *OGLE* towards the Bulge.

The number of objects lensing the light from a background source at any time is given by the optical depth, $\tau \propto R_e^2/m$, where

$$R_e \equiv \left[\frac{4Gmx(L-x)}{c^2 L}\right]^{1/2} \qquad (2)$$

is the Einstein radius of a lens of mass $m$ at distance $x$ along the line of sight, with $L$ the distance to the source star. Towards the LMC and Galactic Bulge, $\tau \lesssim 10^{-6}$ for the disc and halo and so it is necessary to continuously monitor $10^{6-7}$ stars to give a reasonable chance of detecting an event [11]. Since $R_e \propto m^{1/2}$, $\tau$ is independent of $m$ and is insensitive to the assumed MF.

The characteristic timescale of a lensing event is $t_e = R_e/V_t$, where $V_t$ is the transverse velocity of the lens relative to the line of sight. From eqn (2) one sees that the average timescale $\langle t_e \rangle$ scales as $m^{1/2}$, where $\langle t_e \rangle$ is averaged over $x$, $L$ and $V_t$, and so $\langle t_e \rangle$ is sensitive to the assumed MF. The event rate $\Gamma = \tau/\langle t_e \rangle \propto m^{-1/2}$ is also sensitive to the MF. Hence, in principle, the distribution $d\Gamma/dt_e$ provides a powerful measure of the MF.

Since the microlensing technique cannot uniquely determine $x$ and $V_t$ for a particular lens, one can only obtain a crude estimate of individual lensing masses by integrating over all combinations of $x$, $L$ and $V_t$ consistent with the observed timescale $t_e$ and assuming that *a priori* all masses are equally probable [15]. The result is shown in Fig. (1) for the case of a halo event with a timescale of 20 days (typical of the data) observed towards the LMC. The problem is that this estimate itself *depends* on the assumed MF and so it is not a particularly useful quantity for *determining* it. One must therefore rely on observables such as $\Gamma$ and $t_e$.

However, in the case of the halo, $\Gamma$ and $t_e$ also depend on other unknown or poorly constrained parameters, such as the amount of halo flattening, the halo core radius and the halo velocity distribution. For the disc, the relative contribution of the 'thin' and 'thick' components is not well determined. Furthermore, a major difficulty in reconstructing the separate MFs of the halo and disc dark matter arises from their unknown relative contribution to the observed lensing rate. The contributions to lensing from the LMC and the Bulge themselves also need to be determined [16,17]. The angular dependence of $\tau$ should help to constrain at least some of these parameters, although $\tau$ contains no information on the lens velocity distribution. Hence it may only be possible to obtain information on the shape of the MF for one of the components, and even then only provided it dominates the observed lensing rate.

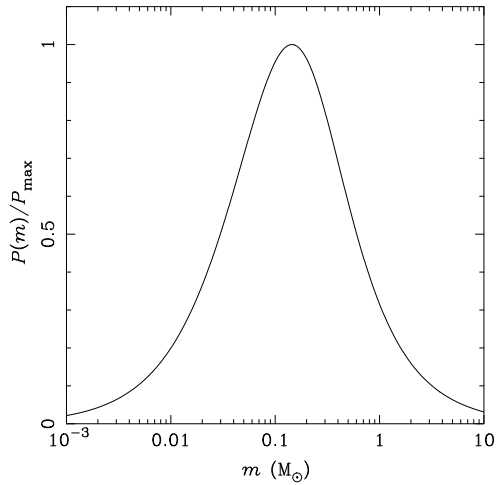

Figure 1. The relative likelihood that a halo lens of mass $m$ produces an observed event of 20 d duration towards the LMC. For other timescales, scale the $m$ axis by $(t_e/20\text{ d})^2$.

Whilst it may be difficult to determine the MF if events *are* detected, one can clearly constrain $\gamma$ if *no* events are detected over some mass range. The EROS CCD search, which is sensitive to low mass objects, failed to find any candidates from its first year of data, which implies $m_l > 10^{-4}$ M$_\odot$ if $\gamma < 2$ for the halo MF, assuming the halo is completely baryonic [10].

## 4. DIRECT SEARCHES

Recent HST observations of a high-latitude field found 5 stars with $2 < V - I < 3$ below $I = 23$ and none with $V - I > 3$ [9]. This is much lower than expected if M-dwarfs comprise either the halo or disc dark matter and so limits their contribution to $< 5\%$ for the halo and $< 15\%$ for the disc at the 95% confidence level. Hence these constraints imply $m_u < 0.1$ M$_\odot$ if $\gamma > 2$ for halo or disc dark matter MFs.

Brown dwarfs are much more difficult to detect and are observable only as very faint infrared sources. To date no 'gold-plated' brown dwarf candidate has been observed. The IRAS point-source survey limits the contribution of 0.08-M$_\odot$ brown dwarfs (the most massive) to less than 50% of the disc dark matter at the 95% confidence level [18] (assuming a local disc dark matter density of 0.15 M$_\odot$ pc$^{-3}$ [19]), consistent with limits based on extrapolating the visible stellar MF. IRAS does not place limits on their contribution to halo dark matter [18].

Future infrared missions such as ISO and SIRTF should improve on IRAS [Fig. (2)], as should the 2MASS ground-based survey. The SIRTF satellite should be capable of detecting brown dwarfs about an order of magnitude fainter than those detectable with ISO. The completed 2MASS all-sky survey should be capable of detecting disc brown dwarfs with masses greater than 0.01 M$_\odot$ or halo brown dwarfs more massive than 0.03 M$_\odot$. There is little prospect that brown dwarfs less massive than 0.01 M$_\odot$ will be detected in the near future using direct searches.

## 5. CONCLUSIONS

Microlensing and direct searches in the optical and infrared represent powerful techniques for detecting compact baryonic matter. Current limits from microlensing already imply that $m_l > 10^{-4}$ M$_\odot$ if $\gamma < 2$ for halo MF and HST observations require $m_u < 0.1$ M$_\odot$ if $\gamma > 2$ for halo or disc MFs, leaving only the brown dwarf mass range ($10^{-3}$ M$_\odot$ – 0.08 M$_\odot$) presently unconstrained [20]. On the other hand, infrared searches over the next decade or so should either detect brown dwarfs more massive than 0.01 $M_\odot$ or place limits on their density. Together with spectroscopic and velocity dispersion data, these searches should allow determinations of the local halo and disc dark matter MFs. It is unlikely that brown dwarfs less massive than 0.01 $M_\odot$ will be directly detected in the near future.

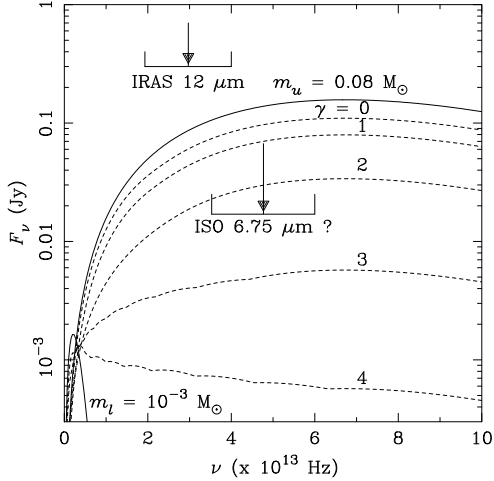

Figure 2. The expected flux from the brightest halo brown dwarf, assuming a local halo density of 0.01 $M_\odot$ pc$^{-3}$. If all the brown dwarfs have mass $m_u$ or $m_l$ (solid lines), the brightest brown dwarf will be the nearest, but if the MF is a power law (dashed lines) then different masses will dominate the flux at different frequencies. The 'effective' sensitivity of the IRAS and ISO satellites are also shown. (The ISO $3\sigma$ limit assumes an integration time of 100 s over an area of 16 deg$^2$, corresponding to a total observing time of 7.5 days.)

The microlensing technique has the added advantages that it is, in principle, sensitive to the entire proposed baryonic mass range from $10^{-7}$ $M_\odot$ – $10^6$ $M_\odot$ and that MF determinations from microlensing are global rather than local (since microlensing is sensitive to objects anywhere along the line of sight). However, in practise, it may prove difficult to reconstruct both the halo and disc MFs from observations. Proposed space-bourne microlensing searches, which together with observations from the ground would provide parallax information [21], will be extremely useful in this regard.

## ACKNOWLEDGMENTS

EJK is grateful to PPARC and the Royal Swedish Academy of Sciences for their financial support.

## REFERENCES

1. J. Binney and S. Tremaine, Galactic Dynamics, Princeton University Press, Princeton, 1987.
2. B. J. Carr, ARA&A 32 (1994) 531.
3. M. Persic and P. Salucci, MNRAS 258 (1992) 14p.
4. T. P. Walker, G. Steigman, D. N. Schramm, K. A. Olive and H. S. Kang, ApJ 376 (1991) 51.
5. G. Efstathiou, J. R. Bond and S. D. M. White, MNRAS 258 (1992) 1.
6. J. Scalo, Fundam. Cosmic Phys. 11 (1986) 1.
7. P. Kroupa, C. A. Tout and G. Gilmore, MNRAS 262 (1993) 545.
8. H. B. Richer and G. G. Fahlman, Nature 358 (1992) 383.
9. J. N. Bahcall, C. Flynn, A. Gould and S. Kirhakos, ApJ 435 (1994) L51
10. R. Ansari, these proceedings.
11. B. Paczyński, ApJ 304 (1986) 1.
12. C. Alcock et al., Nature 365 (1993) 621.
13. E. Aubourg et al., Nature 365 (1993) 623.
14. A. Udalski et al., Acta Astron. 43 (1993) 289.
15. E. J. Kerins, MNRAS submitted.
16. K. C. Sahu, Nature 370 (1994) 275.
17. M. Kiraga and B. Paczyński, ApJ 430 (1994) L101.
18. E. J. Kerins and B. J. Carr, MNRAS 266 (1994) 775.
19. J. N. Bahcall, C. Flynn and A. Gould, ApJ 389 (1992) 234
20. B. J. Carr and E. J. Kerins, in preparation
21. A. Gould, ApJ 421 (1994) L75.